# Tailoring exciton dynamics by elastic strain-gradient in semiconductors

Xuewen Fu[1*], Cong Su[2,4*], Qiang Fu[1], Xinli Zhu[1], Rui Zhu[1], Chuanpu Liu[1], Jun Xu[1], Ji Feng[2†], Ju Li[3‡] and Dapeng Yu[1§]

[1] *State Key Laboratory for Mesoscopic Physics, and Electron Microscopy Laboratory, School of Physics, Peking University, Beijing 100871, China*

[2] *International Center for Quantum Materials, School of Physics, Peking University, Beijing 100871, China*

[3] *Department of Nuclear Science and Engineering and Department of Materials Science and Engineering, Massachusetts Institute of Technology, Cambridge, Massachusetts 02139, USA*

[4] *Yuanpei College, Peking University, Beijing 100871, China*

## ABSTRACT

As device miniaturization approaches the atomic limit, it becomes highly desirable to exploit novel paradigms for tailoring electronic structures and carrier dynamics in materials. Elastic strain can in principle be applied to achieve reversible and fast control of such properties, but it remains a great challenge to create and utilize precisely controlled inhomogeneous deformation in semiconductors. Here, we take a combined experimental and theoretical approach to demonstrate that elastic strain-gradient can be created controllably and reversibly in ZnO micro/nanowires. In particular, we show that the inhomogeneous strain distribution creates an effective field that fundamentally alters the dynamics of the neutral excitons. As the basic principles behind these results are quite generic and applicable to most semiconductors, this work points to a novel route to a wide range of applications in electronics, optoelectronics, and photochemistry.

I. **Introduction**

Precise modulation of electronic structure and electrodynamics inside materials is an essential element for improving the physical and chemical properties of semiconductors, which has been a key theme in applications such as photovoltaics[1-3], photodetection[4], photocatalysis[5] and field-effect transistors[6,7]. Elastic strain engineering can be an effective route for tuning material properties reversibly and continuously. Because strain is $(n+1)n/2$

---

dimensional (where $n$ is effective dimension of the material) continuous field variable that in principle can be precisely controlled, it offers extremely rich possibilities for creating unprecedented materials properties. Of particular note, it has been recently proposed that an inhomogeneous strain field created in a semiconductor atomic membrane can induce spatial concentration of quasiparticles (charge carriers or neutral excitons)[8]. As the characteristic dimension of a material enters into the microscopic regime, such as in micro/nanometre scale sheets and wires, not only do they become mechanically flexible, but they also acquire unusual mechanical strength. We call such materials "ultra-strength materials", as they can withstand enormous elastic deformation up to a significant fraction of their ideal strength, prior to eventual inelastic relaxation by plasticity or fracture[9-11]. Many materials with reduced characteristic dimensions exhibit such unusual mechanical strength. Notable examples are graphene and monolayer molybdenum disulphide ($MoS_2$), which can sustain in-plane elastic strain up to 25%[10,12,13] and 11%[14], respectively.

We also note that nanowires show ultra-strength[9]. Silicon nanowires can sustain 14.3% elastic bending strain without cracking[15], and the ZnO nanowires can be elastically stretched to 7.3% with a red-shift of the energy bandgap up to 110 meV[16]. The large elastic strain limit makes it possible to modulate the physical and chemical properties dramatically via variable elastic strain in the ultra-strength micro/nano-structure semiconductors, as recent experimental progresses have provided evidence for this possibility: (a) significant red-shift and broadening of the near-band-edge (NBE) emission in bent ZnO[17-21] and CdS[22] micro/nanowires, as well as in the uniaxial strained ZnO[16] and GaAs[23] nanowires, have been observed by cathodoluminescence (CL) or photoluminescence (PL) studies; (b) strain-based silicon technology has been widely utilized commercially for enhancing carrier mobility in CMOS transistors[24]; (c) bending deformation in ZnO micro/nanowires has given rise to novel nanogenerators[25,26], piezotronic devices[27] and enhanced light emission in diodes[28]. In all these examples, an exceptionally high elastic strain is present, which greatly "magnifies" the strain effect as compared to its bulk counterparts, and is expected to play a critical role.

Even though elastic strain engineering has already been a standard industry technology in the semiconductor micro/nano-devices design[29], tuning the electrodynamics by elastic inhomogeneous strain has yet to be established experimentally, over sixty years after the initial conception of the notion of deformation potential[30-32]. There are three principal difficulties: lack of high mechanical strength materials, difficulty in precisely creating large inhomogeneous strain, and a proper experimental design for detecting these effects. Recently, Feng et al. theoretically illustrate that it is a feasible implement by employing the elastic strain engineering for generating a continuously varying bandgap configuration in atomically thin semiconducting membranes and spatially concentrating the photo-generated carries. In this paper, by combining theoretical and experimental approaches, we illustrate that by imposing an elastic strain gradient we can tailor the electrodynamics of excitons in semiconducting ZnO microwires (MWs). The key role of an inhomogeneous elastic strain distribution is to give rise to a continuously varying electronic band structures profile within the bent semiconductors micro/nanowires. This is an initial demonstration of the paradigm of elastic strain engineering for tailoring carrier dynamics, which is based on a simple principle and rather generic. As the method for creating the strain gradient here is fairly simple, precise, tunable and versatile, the concept we put forth herein applies to semiconducting materials in general, and may lead to a wide range of applications, such as exciton condensation, broad-spectrum solar energy harvesting and tunable microcavity lasing.

## II. Methods

1. ZnO microwires growth: The synthesis of ZnO microwires was carried out in a horizontal quartz tube furnace by chemical vapor deposition (CVD). The mixture of pure zinc oxide (99.9999%) and graphite powder (molar ratio of 1:1) were loaded in an alumina boat. Sapphire chips with (110) orientation were placed above the source powder as the collecting substrates. The boat was then placed at the center of the quartz tube and inserted into a rapid heating furnace. The system was purged of contaminants with argon gas for more than ten minutes. After this, the growth carrier gas argon was maintained at 200 sccm flow. And the furnace was heated up to 1050 °C in 20 min and then the oxygen (3.0 sccm) was introduced as the reactive gas. The reaction proceeded for 30 min, after which the system was cooled down to room temperature naturally and the substrate was covered by a layer of wax-like products.

2. DFT calculation: Our calculations are based on the density functional theory (DFT) within the generalized gradient approximation (GGA), in the form of Perdew-Burke-Ernzerhof's exchange correlation functional[33,34]. All the calculations are performed using the Vienna Ab-initio Simulation Package (VASP)[35]. Periodic boundary conditions were employed. Geometrical optimizations of ZnO structure are performed until the Hellmann-Feynman forces on the ions are less than $1.0 \times 10^{-4}$ eV/angstrom. The plane-wave basis is used, with a cut-off of 700 eV that converges the total energy to 1 meV/atom. The Brillouin zone is sampled using $8 \times 8 \times 8$ Monkhorst-Pack **k**-point scheme[36].

3. Electrodynamics simulation: The 2D partial differential equation is solved by finite element method implemented in Comsol Multiphysics. Parameters in equation are set to be: the mobility of exciton in ZnO 200 $cm^2\ s^{-1}$, temperature 5.5 K. The lifetime of exciton varies in terms of the bandgap, which in our system ranges from 160 ps (belly side) to 280 ps (back side). The PL pulse in time-dependent simulation is a Gaussian distribution with the time evolution, of which the standard deviation is set to 0.1 ps. Non-flux boundary conditions of belly and back side of the microwire are used, while symmetrical boundary condition is imposed in other two boundaries.

4. Standard 4PB and 3PB ZnO microwires tests: A series SU8 (SU-8, 2015) pillars array, with the diameter of 6.0 um and height of 15.0 um, respectively, were fabricated on a Si substrate by photolithography and developing technology. First, spin coat the SU8 photoresist on the Si substrate (500rpm for 10s and then 3000rpm for 60s). Second, heat the SU8 photoresist (65°C for 1min and then 95°C for 3min). Third, the wafer was exposed under the UV lamp (12 s, 13 mW/cm$^2$) with a pre-designed graph as the mask template. Finally, the wafer was developed by developer solution. After that, micromanipulation was utilized to transfer an individual ZnO microwire with diameter larger than 1.0 μm (much more larger than the exciton diffusion length of ZnO at 5.5K (~100 nm)[37]) and the length over hundreds of micrometers of interesting from the growth substrate on to the prearranged substrate under an optical microscope by using two needle-shaped glass tips. The ZnO microwire was manipulated to suspend over the substrate by sticking in the middle of four SU8 pillars, which restrict the wire in a curved shape after removing the two glass tips. According to ASTM E855-08, in the standard 4PB setup, the bending part between the two inner SU8 pillars is expected to be under a pure bending strain state. Such suspended bending deformation is elastic and the curved micro-wire can resume its original straight state once it is taken out of the SU8 pillars. The standard 3PB measurements were carried out by similar processes.

5. CL measurements: To obtain the optimum spatial resolution with best signal-to-noise ratio, an electron beam (with beam current of about 0.353 nA) was accelerated at 10 KV voltage, which results in the effective interaction of electron beam in ZnO ranging about 100 nm (with 90% power in this region, as supported by Monte Carlo simulation[38]). The CL spectra were carefully collected step by step along the radial direction across the diameter of the ZnO microwires from outer side to inner side by CL spectroscopy (Gatan monocle 3+) at liquid helium temperature (5.5K). The CL spectra were recorded by CCD (Charge Coupled Device) with a scanning range of 300 to 450 nm with spectral resolution of about 0.5 nm. The line-scanning step size was set at about 70~100 nm.

## III. Strain-dependent electronic structures of ZnO MW

Generically, elastic strain corresponds to a reversible change in the arrangement of atoms in a material, which, in the strongly-bound electron limit, causes a fundamental change in the bonding. The change in bonding is immediately linked with change in the electronic quasiparticle energies, hinting at the possibility of strain engineering of quasiparticle dynamics.[8,30] By imposing pure bending deformation on a ZnO MW, we can precisely create an inhomogeneous strain field with a uniform radial strain-gradient. Ideally, the wire develops tangential strain along the hexagonal axis (0001) of the crystal, which varies linearly in the radial direction. For the purely bent MW, the strain is compressive on the belly (inner side), and tensile on the back (outer side), separated by an intermediate neutral plane, as schematically shown in **Fig. 1a**. Previous computational and experimental results indicate that, in ZnO crystals, tensile strain along (0001)-axis reduces the bandgap linearly, while compressive strain widens the gap in an anti-symmetrical manner[18,39-41]. Therefore, the electronic band energies and bandgap are expected to show a continuous spatial and nearly linear variation from belly to back in a purely bent ZnO MW.

In order to obtain the accurate electronic band energy profile at the cross section of the purely bent ZnO MW, we performed density functional theory calculations on the electronic energy band of different strained (*c*-axis) ZnO crystal (see **Methods**). The strain-dependent distributions of the electronic energy bands along the cross section of a purely bent ZnO MW (1.7 μm in diameter) is shown in **Fig. 1b**. Both the valence band maximum (VBM) and conduction band minimum (CBM) linearly shift downwards from the belly side towards back side (see **Fig. 1b**), where CBM is more sensitive than VBM to elastic strain, and the deformation potentials of CBM and VBM are $k_C$ = -52.7 meV/1% and $k_V$ = -21.3 meV/1%, respectively. Correspondingly, the Kohn-Sham quasiparticle gap decreases linearly along radial direction from the belly to the back of the wire as: $\Delta E_{gap} = (k_C - k_V)\varepsilon$ eV, as shown in **Fig. 1c**. It is the elastic strain-gradient in purely bent ZnO MW that gives rise to an effective built-in potential of quasiparticles, which is a universal feature of all non-uniformly deformed semiconductors. It should be noted that the distribution of bandgap will not be affected by the Femi-level pinning effect[42] since conduction and valence bands are always shifted equally with electrostatic potential (see **Fig. 1d**). Therefore, we anticipate a red-shift of optical emission from the back side of pure bent MW, and correspondingly, a blue-shift from the belly side.

### III. Theoretical model and simulation results

In our simulation, we assume that the excitons would not split into free electrons and holes in the strain-gradient field, since the exciton binding energy is strong in ZnO (~60 meV[41]). This assumption will be justified by experimental exciton emission spectra and absence of photo-current, described in the Discussions. The key notion here is that the excitons can migrate towards the tensile side with lower exciton energy before recombination when the electron-hole pairs are generated by steady-state optical or electron beam irradiation, as shown by the schematic graph of the exciton transport mechanism along the radial direction of a bent ZnO MW in **Fig. 1d**.

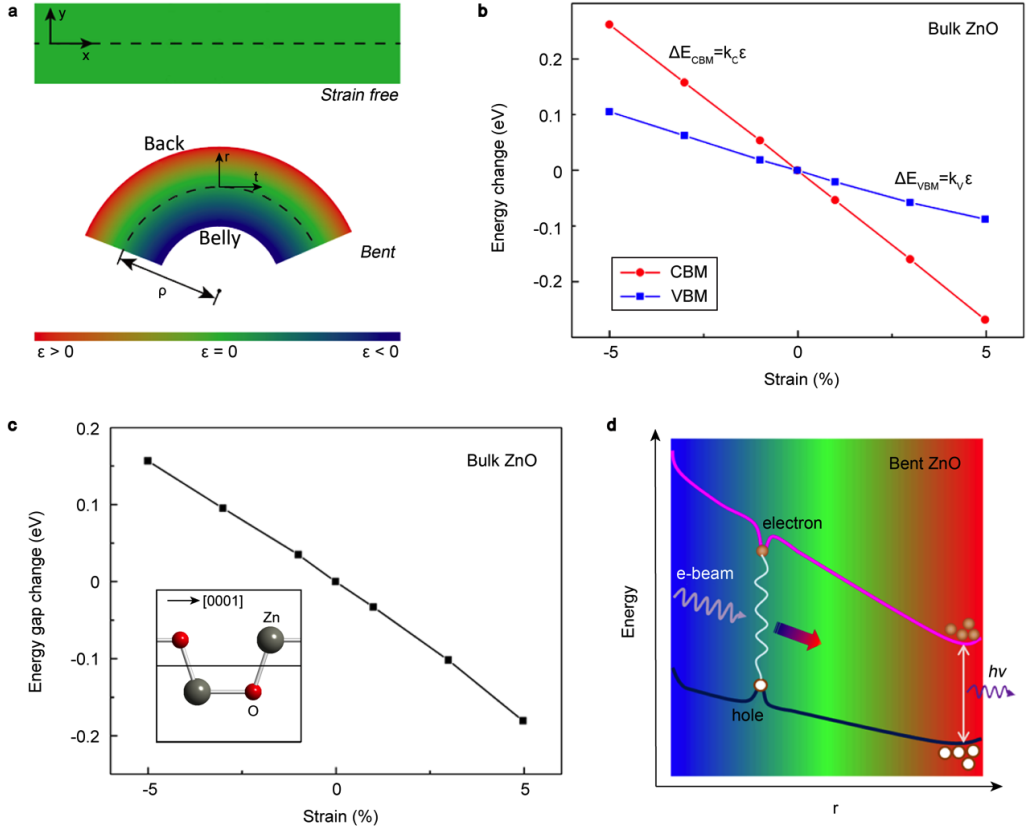

**Fig. 1** Introduction of pure bending strain and strain-dependent electronic and optical properties of ZnO microwire in pure bending deformation. **a,** A schematic configuration of a strain-free (upper) and a pure bent (lower) ZnO microwire. The radius of curvature of the pure bent microwire is denoted as ρ in the graph. **b,** The energy change of conduction band minimum (CBM) and valence band maximum (VBM) as a function of the c-axis strain in bulk ZnO. The slope of CBM is $k_C$ = -52.7 meV/1% and that of VBM is $k_V$ = -21.3 meV/1%. **c,** The energy bandgap change as a function of the c-axis strain in bulk ZnO. Inset: the primitive cell of a bulk ZnO. **d,** The transport mechanism of excited states (excitons) in the cross section of pure bent ZnO microwire. The red and blue colors in Fig. 1a and 1d represent the local tensile and compressive strain, respectively.

In order to reveal the dynamics of excitons in the purely bent ZnO MW, we propose a two-dimensional phenomenological model to simulate the carrier density distribution in a generic MW under pure bending. Note that we use the focused electron beam with high spatial resolution as the excitation source in our simulation. Under steady electron beam irradiation in one confined region of the MW, the continuity equation for the quasiparticles (strongly bound excitons here) reads

$$\partial_t n(\mathbf{r}t) + \nabla \cdot \mathbf{j}(\mathbf{r}t) = g(\mathbf{r}t) - \tau^{-1} n(\mathbf{r}t) \tag{1}$$

where $n(\mathbf{r}t)$ and $\mathbf{j}(\mathbf{r}t)$ are the *number* density and flux of the excitons,

respectively; $g(\mathbf{r}t)$ is the generation rate; $\tau$ is the recombination lifetime of the excitons, taking into account various relaxation pathways. The flux $\mathbf{j}(\mathbf{r},t)$ is

$$\mathbf{j} = -D\nabla n - \mu n \nabla \phi_{\text{e.s.}} \qquad (2)$$

Here the first and second terms correspond, respectively, to contributions from the diffusion of excitons, and the drift current induced by an effective potential gradient; $D$ is the exciton diffusion coefficient; $\mu$ is the mobility of the excitons; $\phi_{\text{e.s.}}(\mathbf{r})$ is the local effective quasiparticle potential arising from the elastic strain within a ZnO MW (see Discussions and Supplementary Materials sections for more details). According to the Einstein relation connecting diffusivity with mobility, $\mu = De/k_\text{B}T$.

The simulation results of the excitons distribution at the cross section in a purely bent ZnO MW (1.7 μm in diameter) with the elastic strain-gradient (defined as the inverse of the radius of bending curvature) of 1.17% μm$^{-1}$ are presented in **Fig. 2**. **Fig. 2a** and **Fig. 2b** show the side and top views of excitons distribution in the pure bending cross section with different electron exciting locations from tensile side to compressive side, respectively. When the incident electron beam focused at the tensile side of the MW with greater tension ($d_s$ = 0.8 μm and 0.4 μm), nearly all the excitons stay localized. As the foci of electron beam move toward the compressive side, we see clearly that the exciton density still peaks at the tensile edge, leaving a streak toward the electron incident point ($d_s$=-0.4 μm and -0.8 μm). Since ZnO is a direct bandgap semiconductor, the radiative recombination rate is proportional to the exciton concentrations. The luminescence at the belly and back of the bent wire is blue- and red-shifted, as expected. Therefore if we collect the luminescence spectra excited by focused electron beam from back to belly along the pure bending cross section, the spectra should exhibit a dominant constantly red-shifted emission peak, and a weak broad emission band with higher photon energy, as schematically shown in **Fig. 2c**.

The inherent assumption in the model above is that the exciton binding energy, though strong (60 meV[41]), is still much smaller than the band gap (above 3 eV), therefore the variation in exciton binding energy due to elastic strain (as well as Coulomb field polarization) is regarded as insignificant compared to the contribution due to the band gap change due to elastic strain (a point to be returned to in the Discussions). For simplicity we have ignored the 3D nature of the MW geometry and treated it as if it was a 2D problem.

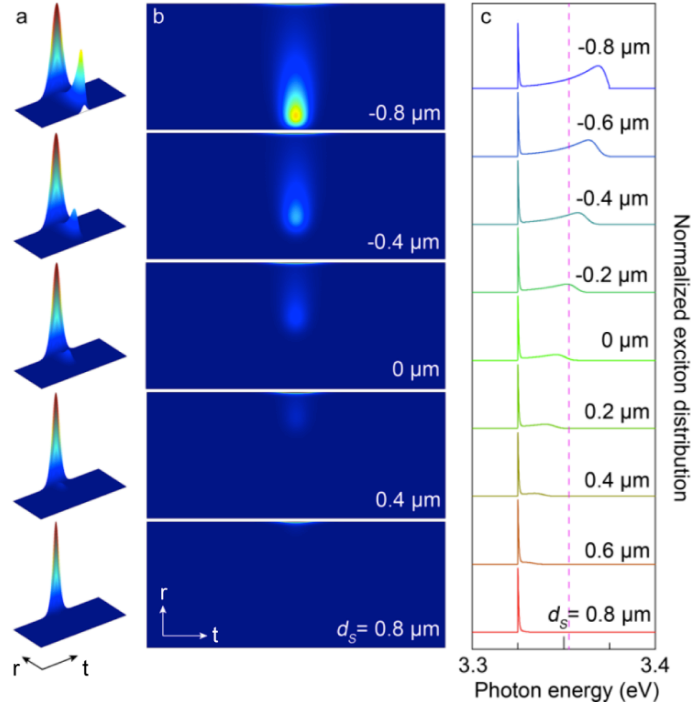

**Fig. 2** Simulation results of the excited states (excitons) distribution in the cross section of a pure bent ZnO microwire with elastic strain-gradient of $g=1.17\ \%\ \mu m^{-1}$. **a,** Three-dimensional views of excited states distribution at the cross section of the pure bent ZnO microwire with different electron beam exciting locations from tensile side to compressive side (bottom to up). x direction is set along c-axis of the pure bent microwire, and the direction perpendicular to the x-y plane represents the normalized concentration of excited states. **b,** Top views of excited states distribution excited states distribution at the cross section of the pure bent ZnO microwire with different electron beam exciting locations from tensile side to compressive side (bottom to up). The electron beam exciting location with respect to the neutral surface is denoted as $d_S$ (similar hereinafter). The width of microwire is 1.7 μm, where the top and bottom boundaries of pure bent microwire are at the places of $d_S = 0.85$ μm and -0.85 μm respectively. **c,** The normalized excited states distributions of the pure bent ZnO microwire as a function of energy bandgap with electron beam exciting location moving from tensile side to compressive side (bottom to up). The pink dashed line indicates the original strain-free position of the NBE emission photon energy in ZnO microwire.

## IV. Experimental design and results

To confirm our theoretical analysis, we carry out high spatial and energy resolution CL characterization on ZnO MWs under pure bending at 5.5 K, with a standard four-point-bending (4PB) setup (see Method Section). **Fig. 3c** shows the typical top-view SEM image of a ZnO MW (2.8 μm in diameter) in such 4PB setup.

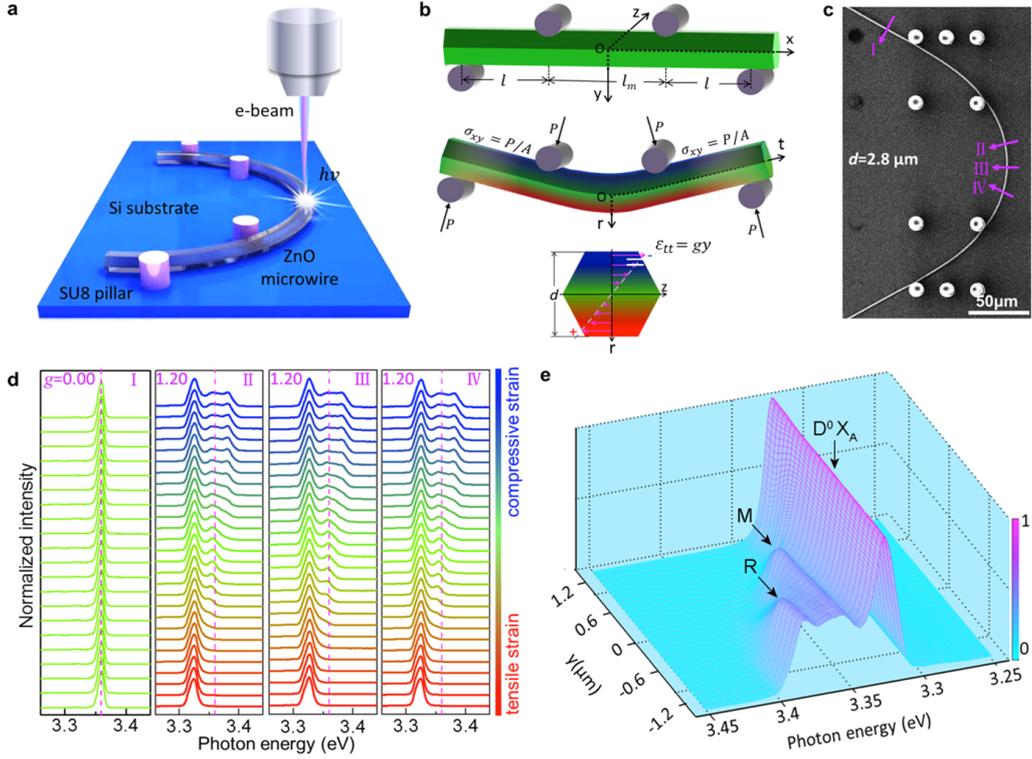

**Fig. 3** Schematic diagram for CL measurement and elastic pure bending strain distribution in a standard 4PB ZnO microwire. **a**, Schematic diagram for line-scanning CL measurements on the pure bending region of a 4PB ZnO microwire step-by-step from tensile to compressive edges. **b**, Distribution of strain $\varepsilon_{tt}$ in a standard 4PB wire subjected to force $P$ at the four points. The red and blue colors illustrate the tensile and compressive regions in the bent wire, respectively. The figure bottom insert shows the c-axis strain $\varepsilon_{tt}$ distribution at the pure bending cross section. **c**, Top-view SEM image of a typical 4PB ZnO microwire with diameter of $d$=2.8 μm. **d**, Line-scanning CL spectra on the strain-free section "I" and from the pure bending region ("II"-"IV") with constant strain-gradient $g$=1.20 % μm$^{-1}$ at 5.5 K. The red and blue colors in the CL spectra represent the tensile and compressive strain, respectively. The pink dash lines indicate the original strain-free position of the NBE emission peak. **e**, Three-dimensional graph of the NBE emission spectra (normalized in intensity) along the cross-section "III" of the 4PB ZnO microwire.

**Fig. 3b** schematically illustrates the strain distribution in a ZnO wire in such a 4PB setup, in which the four pillars divide the wire into five segments along the wire axis. The middle segment is in a pure bending state, with the c-axis strain distributing linearly from inner side to outer side as $\varepsilon_{tt} = r/R\,(|r| \leq d/2)$ where $d$ is the diameter of the wire, and $R$ the radius of curvature of the local neutral plane. We have adopted a local $r$-$t$ coordinates (See **Fig. 3b**), where $r$ and $t$ are the mutually perpendicular local radial and tangential directions ($r$ is measured from the neutral plane).

**Fig. 3a** schematically shows the line-scanning CL measurement procedure. In these measurements, the electrons are injected step by step along a cross section in the pure bending region of the ZnO MW. From all the data reported herein, the ZnO MWs are elastically deformed, as these wires fully restores the initial straight geometry upon release of the constraints.

The CL spectra measurements are performed at 5.5 K in a liquid helium flow cryostat, under which only one intensive NBE peak, corresponding to a neutral donor bound exciton $D^0X_A$ [43] [44] is dominant around 3.359 eV for strain-free ZnO wires (see **Fig. S2**). **Fig. 3d** shows four sets of CL spectra, each corresponding to a cross section of the ZnO MW. One set of spectra are collected for a cross section in the straight region ("I" **Fig. 3c**). Three other sets of spectra correspond to three cross sections in the purely bent region ("II", "III" and "IV", respectively). Clearly, the line-scanning CL spectra at "II", "III" and "IV" are identical, confirming that the 4PB wire is indeed in the elastic pure bending state where the fraction between the substrate and ZnO MW could be ignored. The line-scanning CL spectra collected from the cross section in the bent region of the wire are still dominated by the $D^0X_A$ luminescence peak, but are clearly red-shifted compared with the emission peak in the strain-free region by 33 meV. A key observation here is that two weaker but identifiable peaks (labelled as M and R) with higher energies emerge with the electron injection spot moving towards the inner side. The energy of M peak stays almost unchanged, while the energy of R peak slightly increases.

This experimental result is in good qualitative agreement with the theoretical simulation presented in **Fig. 2**. The only major difference between experiment and simulation is that, there are three peaks (one intensive peak and two weak peaks) when electron beam is focused on the belly side of the bent wire, whereas only two peaks (one intensive peak and one weak peak) are found in the simulation results. This additional peak M in experimental spectra can be attributed to one of the whispering gallery mode (WGM) resonances that arise from the light internal reflection at the side facets within the hexagonal cross section of the MW.[45] It is a completely photonic characteristic and the position of the WGM is determined by the refractive index and diameter of the cross section. The assignment of peak M to the WGM is supported by two observations: (1) the peak stays almost unchanged for the same wire regardless of the position of the electron beam irradiation; and (2) the peak M changes with the diameter of the wire according to resonance condition of the WGM.

To further reveal the electrodynamics behavior in a 4PB ZnO MW under different pure bending states, we carried out CL measurements on one same ZnO MW under three different strain-gradients: $g_I$=0.00% μm$^{-1}$ ($\varepsilon_{tt}^{max} = 0.00\%$), $g_{II}$=0.57% μm$^{-1}$ ($\varepsilon_{tt}^{max} = 0.48\%$) and $g_{III}$=1.16% μm$^{-1}$ ($\varepsilon_{tt}^{max} = 0.98\%$), as shown by the 30°-tilted SEM images in **Fig. 4a** from left to right. The line-scanning CL spectra collected from the same cross section under the three different pure bending states indicated by the pink arrows in **Fig. 4a** are presented in **Fig. 4b**. Similar to the line-scanning CL spectra results presented in **Fig. 3d**, only one intensive constant red-shifted $D^0X_A$ dominates the CL spectra at the whole section from tensile to compressive edges under different bending states. When the ZnO MW is bent from the strain-free state "I" into state "II" and state "III", $(D^0X_A)_I$ shifts to $(D^0X_A)_{II}$ and $(D^0X_A)_{III}$, with red shifts of 12.97 meV and 24.43 meV, respectively. It is worth noting that $(D^0X_A)_{II}$ has larger full width at half maximum (FWHM) than $(D^0X_A)_{III}$. This larger peak broadening is easy to be understood by our theoretical model: In state "II", the strain-gradient induced built-in field is not large enough to immediately drive the excitons to the edge of the back-side. Consequently, the luminescence has a broad spectrum corresponding to a broader distribution of excitons at the cross section.

To have a comprehensive view of the NBE emission energy evolution from states "I" to "III", the spectra of all three bending states are displayed in **Fig. 4c**. With the electron excitation spot moving to the compressive edge, the M peak enter a stable region with a nearly constant energy of 3.365 eV, while the R peak shows a gradual blue shift to 3.381 eV (please refer to **Fig. S3** in Supplementary materials for more discussions). In accordance with our simulation result, the energy red shift of $D^0X_A$ peak indeed varies linearly with the increasing bending strain, as shown in **Fig.** $\Delta E = k\varepsilon_{tt}^{max}$ **4d**. The linear fitting yields (without considering Poisson effect), where $k = \partial(\Delta E)/\partial\varepsilon_{tt}^{max} = -2.49 eV$ is the deformation potential of the pure bent ZnO MW. As the pure bending is different from hydrostatic and uniaxial compression, the deformation potential obtained in our experiment is larger than that of Dietrich's experiments (-2.04 eV) [18] and smaller than that of Shan's pressure experiments (-3.92 eV) [40].

The above analysis for 4PB MWs could be generalized to three-point-bending (3PB) MWs with a more complex inhomogeneous strain field. In a 3PB MW, strain-gradient along the *c*-axis direction (t-direction in **Fig. 1a**) does not vanish since the radius of curvature varies along the MW. This strain-gradient would also induce a built-in field, by which the drift of excitons are impelled along the t-direction. The CL measurement of a standard 3PB MW (2.2 μm in diameter) is performed at 5.5K. Except for the relative intensity of the emission peaks, the characteristics of the spectra measured in the bending region show no essential difference with that of 4PB MW (**Figs. S4d** and **S4e**), which are in good agreement with our calculation (**Figs. S4a**, **S4b** and **S4c**). As the strain-gradient along the t-direction is much smaller than that of the radial direction (r-direction), the drifting of excitons along t-direction is too weak to influence the distribution of excitons along the cross section, and hard to be observed in the CL spectra. The different emission intensity of the peaks between 3PB MW and 4PB MW is due to the change of the distribution of the excitons at the belly and back side.

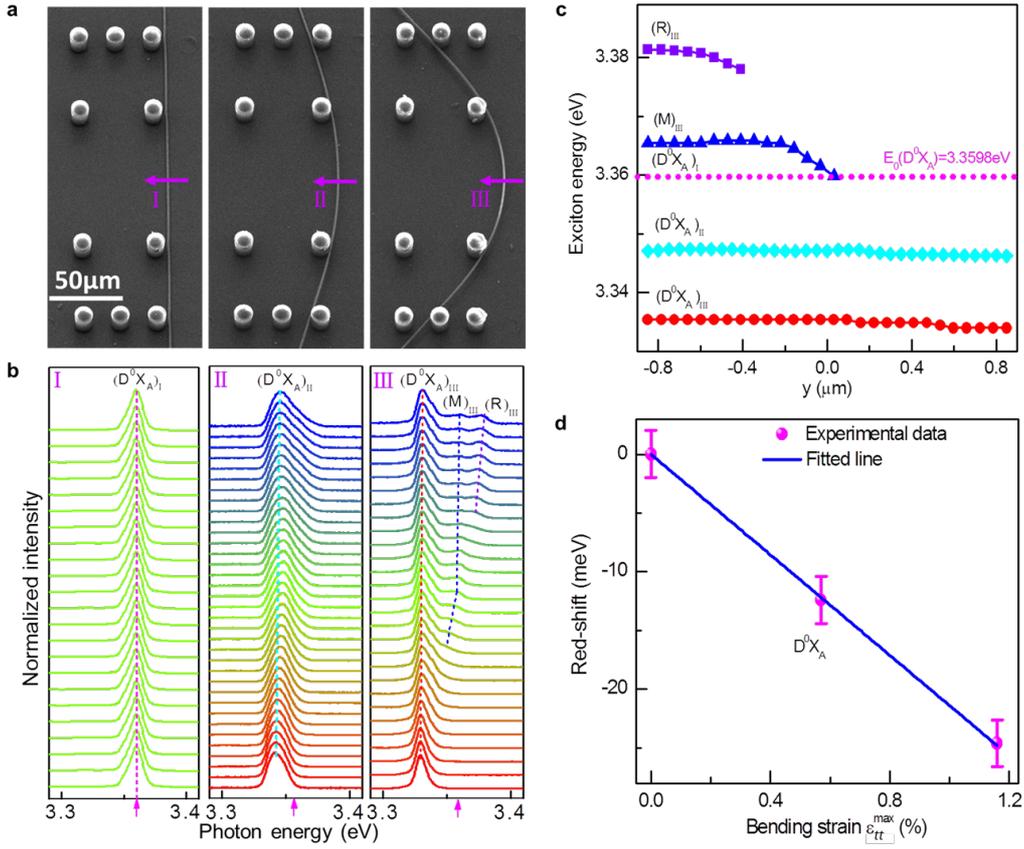

**Fig. 4** Red-shift of the NBE emission photon energy in a 4PB ZnO microwire ($d$=1.7 μm) under different elastic pure bending states at 5.5 K. **a**, $30°$-tilted SEM images of the same individual ZnO microwire in the 4PB setup before bending (left), bent to $g_{II}$=0.57% μm$^{-1}$ (middle) and to $g_{III}$=1.16% μm$^{-1}$ (right). The pink arrows indicate the CL line scanning direction and the same studied cross section. **b**, The collected line-scanning CL spectra from the same cross section under three different pure bending states ("I", "II", "III") shown in Fig. 4a. Under the second pure bending situation with $g_{III}$=1.16% μm$^{-1}$ two weak emission peaks M and R emerge gradually with electron beam exciting spot moving towards the compressive edge. The red and blue colors represent tensile and compressive strain, respectively. The upward pink arrows at the bottom of the spectra indicate the original strain-free position of the NBE emission peak D$^0$X$_A$. **c**, Positions of the NBE emission peaks with electron beam exciting spot moving along the cross-section at different pure bending states. (D$^0$X$_A$)$_I$: D$^0$X$_A$ emission photon energy under the strain free state "I"; (D$^0$X$_A$)$_{II}$: D$^0$X$_A$ emission photon energy under the first pure bending state "II" with $g$=0.57% μm$^{-1}$; (D$^0$X$_A$)$_{III}$: D$^0$X$_A$ emission photon energy under the second pure bending state "III" with $g$=1.16% μm$^{-1}$; (M)$_{III}$ and (R)$_{III}$: emission photon energies of the weak M and R peaks under the second pure bending state "III" with $g$=1.16% μm$^{-1}$. **d**, Linear red-shift of the dominant emission (D$^0$X$_A$) photon energy of the NBE emission against the elastic pure bending strain. The linear fitting process in Fig. 3d yields the bending deformation potential of -2.49 eV for the 4PB ZnO microwire. The error bars of the data are determined by the energy resolution of the CL spectroscopy.

# V. Discussions

One concern for the reliability of our experiment results is the electron beam irradiation effects in the CL measurement. We find that the electron beam irradiation can only weaken the CL spectra intensity slightly during the CL measurement, but does not change the peak energy even with increasing irradiation duration up to 3 minutes. The dependence of ZnO CL spectra on electron beam irradiation duration described in Reynolds' work[46] reported the same results.

An issue that must also be reckoned with is possibility that the piezoelectric polarization induced built-in transverse electric field[47,48] enters into the physics discussed here. In a bent ZnO wire, this bending deformation induced piezoelectric electric field would result in inhomogeneous band bending inside the ZnO MW. Xu et al. attributed the net red-shift of the NBE emission photon energy in bent ZnO wires to the piezoelectric field modification of the photoexcited carriers[21]. But as it amounts to an electrostatic potential which shifts or bends the valence and conduction bands equally, we do not believe the piezoelectric field will significantly alter the frequency of light emitted. When electron-hole pairs are generated in the bending region of the 4PB and 3PB ZnO MWs, the transverse piezoelectric field could influence the motions of separated electrons and holes. However, due to the significant exciton binding energy in ZnO, the generated electron-hole pairs prefer to stay coupled. The center-of-mass force, which drives the motion of a neutral exciton, has vanishing coupling to the piezoelectric field, to the leading order. Fundamentally, if we ignore the polarization effect, the excitons are neutral particles which should not "care" about the Coulomb potential; it should be the strain-induced band gap change, or deformation potential, that these neutral excitons "care" the most about.

Formalistically, a neutral exciton's total energy is expressed as $E_\text{g} - E_\text{B}$. The gap energy $E_\text{g}$ represents the energy difference between single electron injection (electron affinity) and single hole extraction (ionization potential) at well-separated locations in the material. The exciton binding energy $E_\text{B}$ is the energy reduction when the electron and hole are brought into proximity. Since band-bending due to electric field has no effect on the band gap at a given location, we have argued that $E_\text{g}$ is independent of the electric field $\mathbf{E}$, and only cares about the strain tensor $\varepsilon$:

$$E_\text{g} = E_\text{g}^\text{ref.} + \varepsilon : (\partial E_\text{g}/\partial \varepsilon)^\text{ref.} + \mathcal{O}(\varepsilon^2). \tag{3}$$

Here the electric field may be that of the piezoelectric polarization, $\mathbf{E} = \eta \mathbf{P}$, where $\eta$ is the inverse dielectric constant of the material. On the other hand, $E_\text{B}$ does care about $\mathbf{E}$ and $\varepsilon$ :

$$E_{\text{B}} = E_{\text{B}}^{\text{ref.}} + \varepsilon : (\partial E_{\text{B}}/\partial \varepsilon)^{\text{ref.}} - \frac{1}{2}\alpha^{\text{ref.}} : \mathbf{EE} + \mathcal{O}(\varepsilon^2, \varepsilon\mathbf{E}^2), \tag{4}$$

where $\eta$ is the electric polarizability of the exciton, but since $E_{\text{B}} << E_{\text{g}}$ it is unlikely to contribute much. The total driving force $\mathbf{F}$ on exciton center-of-mass motion should therefore be

$$\mathbf{F} = -(\nabla \varepsilon) : (\partial E_{\text{g}}/\partial \varepsilon + \partial E_{\text{B}}/\partial \varepsilon)^{\text{ref.}} + \frac{1}{2}\alpha^{\text{ref.}} : \nabla \mathbf{EE} + \mathcal{O}(\nabla \varepsilon^2, \nabla \varepsilon \mathbf{E}^2) \tag{5}$$

where in the present work we only took the first-order term without the change in the exciton binding,

$$\mathbf{F} = -(\nabla \varepsilon) : (\partial E_{\text{g}}/\partial \varepsilon)^{\text{ref.}} \tag{6}$$

which is expected to be dominant in the drift-diffusion equation (1), (2). Notice that in this general formal analysis, the polarization field only enters as second order terms.

To verify that we are indeed dealing with mainly charge-neutral phenomena that does not couple strongly to the electric field, we perform an experiment to measure the photo-current (bias of 0 V) across a purely bent ZnO wire under UV (325 nm) laser excitation, and found that there is essentially zero photocurrent. This is opposite to the expectation from a piezoelectric mechanism, where the polarization field should give rise to a photo-current.

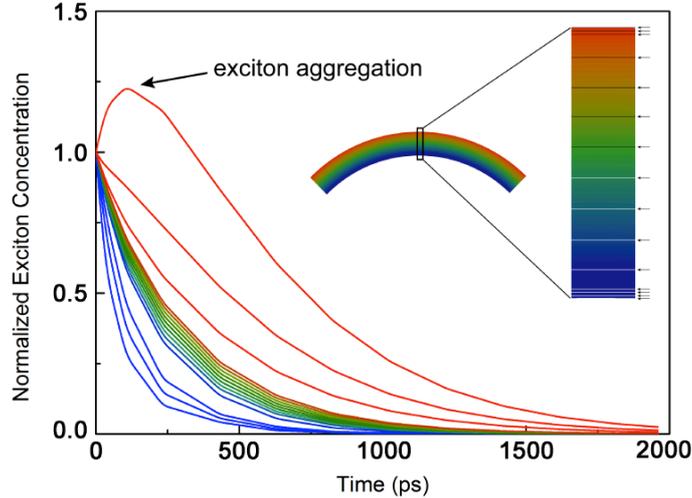

**Fig. 5** The time-dependent simulation result of exciton concentration in ZnO MW under PL excitation. The color of each line, representing the energy of the bound exciton, is in accordance with the strain distribution in MW, as shown by the inset figure. Fifteen groups of excitons with different energies are sampled on ZnO MW marked by lines in the inset color bar. The increase of the concentration near the back side of ZnO MW after t=0 ps indicates the aggregation of exciton under the strain-gradient scheme.

The strain gradient effect proposed herein may also be detected in ultrafast optical measurements. We carry out time-dependent simulations of the dynamical process of excitons in a time-resolved PL experiment. In this experiment, the entire bending cross section of the purely bent ZnO MW is excited by an optical pulse. We depict the simulated decay process of luminescence with the time evolution in **Fig. 5**. The time evolution of luminescence intensity is remarkably different for different energies. The high-energy emission decays super-exponentially. The intensity of the low-energy emission undergoes an initial rise. This is a direct consequence of the exciton migration along an internal energy gradient. Therefore, ultrafast dynamics measurements with spectral and time resolution will be an interesting experiment, to directly reveal the detailed decay kinetics of the excitons in an inhomogeneous strain field.

The experimental evidence for exciton drift in the inhomogeneously strained semiconductor micro/nanowires should be of great potential for exciton condensation, lasing, photovoltaic conversion and optical detection. The unique feature of these devices is that the emission photon energy and intensity, as well as the absorbent light frequency can be continuously tuned by varying the applied bending strain. Since a neutral exciton is a bosonic excitation, the geometric concentration of excitons could also be used to create Bose-Einstein condensate (BEC) of excitons at sufficiently low temperatures, especially near geometric singularities inside the specimen such as surface asperities and cracks under stress. The fabrication of these devices should be very simple, as the inhomogeneous strain in semiconductor micro/nanowires can be easily and precisely tuned by controlling the bending curvature and specimen geometry.

In conclusion, through systematic high spatial and energy resolution cathodoluminescence investigations on purely bent ZnO MWs at 5.5 K, we find that the NBE emission spectra exhibit a dominant red-shift peak at the whole bending cross section and two weak higher energy emission subbands emerge when entering the compressive side. In excellent agreement with our simulations, the experiments demonstrate that dynamics of excitons can be artificially tailored by the elastic strain gradient. The key is the deformation potential-driven drift of the excitons. An effective potential is created by the strain gradient that directs the motion of the excitons. As the model we propose is rather generic, which is in excellent agreement with the spectroscopic measurements, one can reasonably argue that the phenomena described herein apply generally for semiconductors. The ability to tune the exciton dynamics and the directed concentration of excitons may be of important implications for such applications as exciton BEC, micro-cavity exciton lasing, broad band photovoltaic conversion and optical detection.

## Acknowledgements

This work is supported by NSFC (Grant No. 11174009), the National 973 Programs of China (2009CB623703, 2012CB619402, 2013CB921900). The authors are also grateful to the financial support from the Sino-Swiss Science and Technology Cooperation Program (2010DFA01810), and NSFC/RGC (N HKUST615/06). J.L. acknowledges support by NSF DMR-1008104 and DMR-1120901. We thank Profs. Zhonglin Wang and Junren Shi, and Dr Xiaofeng Qian and Mr Ran Duan for useful discussions.